\documentclass[aps,fleqn,showpacs,twocolumn]{revtex4}

\usepackage{graphicx}
\usepackage{amsmath}
\usepackage{amsfonts}
\usepackage{amssymb}

\def\eps{\varepsilon}
\def\del{\delta}

\renewcommand{\imath}{\mathrm{i}}

\newcommand{\bg}[1]{\ensuremath{\mbox{\boldmath$#1$}}}

\def\e#1{\mathrm{e}^{#1}}

\begin{document}
\title{Non-adiabatic Rectification and Current Reversal in Electron Pumps}
\author{Alexander Croy}
\email{croy@chalmers.se}
\affiliation{Department of Applied Physics,
Chalmers University of Technology, S-41296 G\"{o}teborg, Sweden}

\author{Ulf Saalmann}
\affiliation{Max-Planck-Institute for the Physics of Complex Systems,
   N\"{o}thnitzer Str.~38, D-01187 Dresden, Germany}

\date{\today}


\begin{abstract}\noindent
Pumping of electrons through nano-scale devices is one
of the fascinating achievements in the field of nano-science
with a wide range of applications. To optimize the performance of
pumps, operating them at high frequencies is mandatory. We consider
the influence of fast periodic driving on the average charge transferred through 
a quantum dot. We show that it is possible to reverse the average
current by sweeping the driving frequency only. 
In connection with this, we observe a rectification of the average current for high frequencies.
Since both effects are very robust, as corroborated by analytical results for harmonic driving, they offer a new way of controlling electron pumps.
\end{abstract}

\maketitle

The possibility to pump charges or spins through nanoscale devices despite
the absence of a bias voltage \cite{th83,kojo+91}, 
shows the intriguing potential of driven
nano-devices. It has consequently led to an increasing interest in such
systems over the past decades. This interest was partly driven by the prospect
of achieving {\it single-electron} pumping and thus creating a unique nano-scale
single-electron source. However, it was soon realized that electron pumps may also help to close the metrological triangle, because they provide a connection between frequency and current \cite{fl04,ke08}.
These developments were substantially aided by the rapid experimental and technological progress in controlling and fabricating nano-scale devices. In particular,
it recently became possible to realize charge pumping in the GHz regime \cite{blka+07,funi+08,kaka+08,giwr+10}, 
which leads to a significant increase of the pumped current.

The adiabatic limit of charge pumping, where the driving frequency $\Omega$
is much smaller than the typical charging/discharging rate $\Gamma$, is very well
understood \cite{br98}. In this case, one can resort to well-established 
time-independent formalisms to describe the electron transport 
and many theoretical works have considered adiabatic pumping in various systems \cite{br98,zhsp+99,enah+02,mobu01,cago+09}. 
The opposite limit of very fast pumping has also been studied \cite{hawe+01,brbu08}; to large extent in the context of photon-assisted tunneling 
\cite{brsc94,blha+95,ooko+97}. However, the borderland between 
these limits is lacking a comparably systematic understanding.
Here, one is faced with an inherently non-equilibrium problem, while due to the
similar time-scales a perturbative description is not possible.
To address this problem, numerical calculations in the time-domain \cite{mogu+07a,stku+08a}
or methods based on Floquet theory have typically been used \cite{stwi96,mobu02,stha+05,armo06}. 
Only recently, within the diagrammatic real-time transport theory, has a summation
to all orders in $\Omega$ in the limit of weak tunnel-coupling and moderate pumping frequencies been achieved \cite{cago+09}, revealing interesting non-adiabatic effects for spin and charge pumping.

\begin{figure*}
	\includegraphics[width=\textwidth]{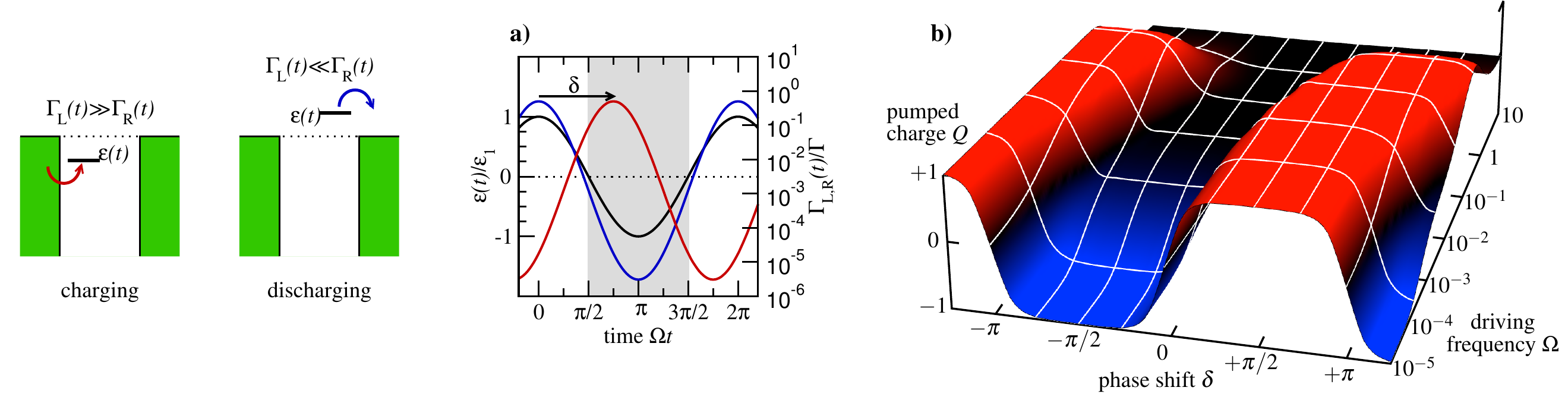}
	\caption{Schematic representation of the pumping device:
	The level $\eps$ and the couplings $\Gamma_{\rm L,R}$ to left and right reservoir oscillate in time,
	 inducing periodic charging and discharging.
	a) Time dependence of $\eps$ (black line, left axis) and $\Gamma_{\rm L,R}$
	 (red and blue lines, right axis) according to Eq.\,\eqref{eq:driving} for a phase delay of $\delta{=}3\pi/4$.  
	b) Pumped charge $Q$ per period as a function of frequency $\Omega$ and phase
		shift $\del$ obtained from an NEGF calculation.
		 The parameters defining the driving and the reservoirs are as follows:
			$\varepsilon_0=0$, $\varepsilon_1=20 \Gamma$,
			 $\eta_\alpha=6$, $\Gamma^0_{\rm L,R} = \Gamma e^{-\eta_\alpha}/2$, 
			$\mu_\alpha = 0$, and $k_{\rm B}T=\Gamma/10$.
		}
		\label{fig:qdpc3D}
\end{figure*}%
\begin{figure}[b]
	\includegraphics[width=\columnwidth]{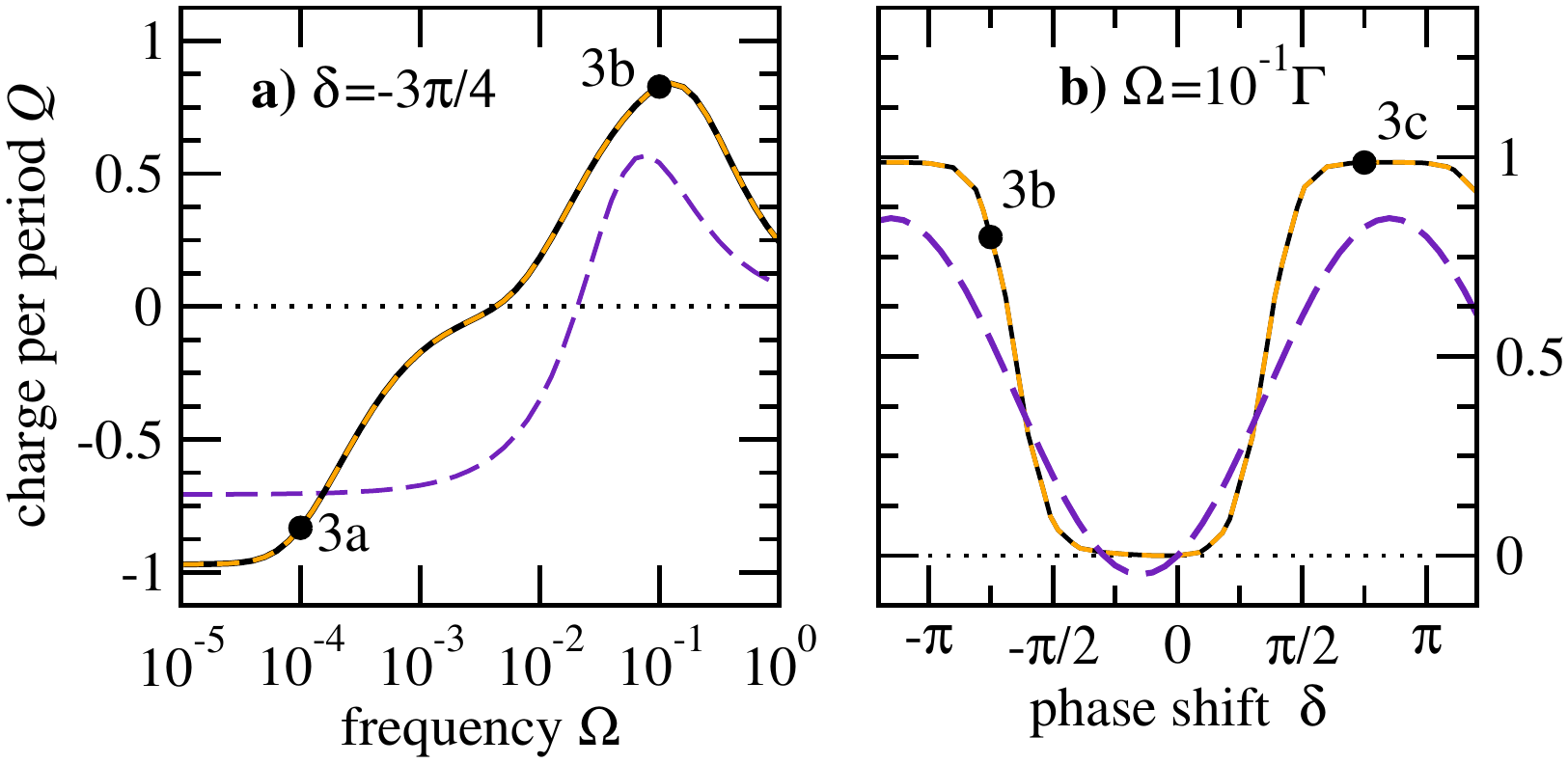}
	\caption{Pumped charge per period $Q$ for the same parameters as in Fig.\,\ref{fig:qdpc3D}. 
	   NEGF (black lines) and rate-equation (broken orange) results are indistinguishable.		
    {\bf a)} Current reversal for the phase shift $\delta=-3\pi/4$.
    {\bf b)} Current rectification for the driving frequency $\Omega=\Gamma/10$.
    Parameters for which the time evolution is shown in Fig.\,\ref{fig:timeev} are marked by black points.
		Result form the harmonic model [Eq.\,\eqref{eq:Qharm} with $\Gamma_{0}=\Gamma_{1}=\Gamma/20$] are shown as indigo/long-dashed line.
		}
		\label{fig:qdpc}
\end{figure}%
In this Letter we demonstrate the drastic consequences of non-adiabatic driving for the pumped charge per period in a generic system and arbitrary frequencies. 
In particular, we show that it is possible to  {\it reverse the average current} by sweeping the driving frequency. 
This reversal is shown to be a hallmark of the borderland between very slow and fast driving regimes. 
It does not rely on interference effects \cite{mobu02} and appears to be robust with respect to different driving schemes.
The reversal effect is associated with a {\it non-adiabatic rectification} of the pumped current for high frequencies.

In our case the electron pump is realized using a quantum dot (QD), which
is coupled via tunnel barriers to source (S) and drain (D) contacts. These are connected to larger electron reservoirs. The total Hamiltonian is $H = H_{\rm dot} + H_{\rm res} + H_{\rm tun}$, where the first term describes the QD itself, $H_\mathrm{\rm dot} = \eps(t) \hat{c}^\dagger \hat{c}$,  the second term characterizes
the attached contacts, $H_\mathrm{\rm res} = \sum_{\alpha \in \mathrm{S},\mathrm{D}} 
\sum_{k s} \eps_{\alpha k} \hat{b}^\dagger_{\alpha k} \hat{b}_{\alpha k}^{}$,
and the third term accounts for
the tunnel coupling between QD and contacts,
$H_\mathrm{\rm tun} = \sum_{\alpha k} \,T_{\alpha k}(t) \hat{b}^\dagger_{\alpha k} \hat{c}^{}
+ \mathrm{h.c}$. Here, $\hat{c}^\dagger$ and $\hat{b}^\dagger_{\alpha k}$ create an electron
in the QD state and in the reservoir state $\alpha k$, respectively.
The transport through the QD is mainly characterized by the tunneling rate
$\Gamma_\alpha(t) = 2\pi \sum_{k} |T_{\alpha k}(t)| \delta(\varepsilon-\eps_{\alpha k})$,
which is given in terms of the time-dependent tunneling amplitudes $T_{\alpha k}(t)$.
In order to calculate the pumped charge per period, we use the framework of NEGFs
\cite{wija+93,jawi+94} in
conjunction with an auxiliary mode expansion \cite{crsa09a}. This method
is very flexible and allows for treating arbitrary time-dependencies of the
parameters entering the Hamiltonian.

Specifically, we use the following time dependence for the QD level and
the couplings to the reservoirs,
\begin{subequations}\label{eq:driving}
\begin{align}
	\varepsilon(t) &= \varepsilon_0 + \varepsilon_1 \cos( \Omega t )\;, \\
	\Gamma_\alpha(t) &= \Gamma^0_\alpha \exp\left[ \eta_\alpha \cos( \Omega t - \delta_\alpha) \right]\;.
\end{align}
\end{subequations}
This choice reflects the experimental situation of modulated voltages, which will in general lead to an exponential dependence of the tunnel couplings on this modulation 
\cite{kaka+08,giwr+10}. The most important aspect of Eqs.\ \eqref{eq:driving} is, however, the 
inclusion of the phase shifts $\delta_\alpha$ allowing for an offset of the coupling oscillations with respect each other. 
As in earlier work \cite{kaka+08}, we use $\del_{\rm L}\equiv\delta$ and $\delta_{\rm R}=0$ in the following. In Fig.\ \ref{fig:qdpc3D}a the time dependence of $\varepsilon$ and $\Gamma_\alpha$ is shown for a specific delay of $\del=3\pi/4$.

The time-dependent driving \eqref{eq:driving} induces currents $J_{\rm L}$ and $J_{\rm R}$ from the left and the right reservoir, respectively.
The net charge, pumped from the left to the right reservoir within one period $\tau\,{\equiv}\,2\pi/\Omega$, can be obtained by the integral
\begin{equation}\label{eq:Qdef}
Q=\frac{1}{2}\int_{0}^{\tau}{\rm d}t'\left[J_{\rm L}(t')-J_{\rm R}(t')\right].
\end{equation}
In numerical calculations, the respective equations are propagated until the charge per period converges.
Figure \ref{fig:qdpc3D} shows the numerical results for the pumped charge $Q$ as a function of frequency $\Omega$ and phase shift $\del$.
For very low frequencies, $\Omega\ll\Gamma$, one finds the expected behavior
of $Q$ in dependence on $\del$: For negative shifts
the pumped charge is positive, while for positive shifts it is negative, which is known as peristaltic pumping \cite{br98}.
In striking contrast, one observes for higher frequencies ($\Omega\gtrsim10^{-2}\Gamma$) that the net current always flows in one direction.  
This implies for a negative phase delay ($\delta{\approx}{-}3\pi/4$) that by sweeping the driving
frequency one can change the sign of the pumped current per period or, in other words, reverse the direction of the average current. 
These effects\,---\,rectification and current reversal\,---\,are the central result of this Letter.

In the following, we analyze the pumping using a simple rate-equation description
of the electron transport, which is valid for $\eps_1\gg\Gamma,k_{\rm B} T$.
As we will show, this description is sufficient to reveal the basic mechanisms behind both effects.
The currents $J_{\rm L,R}$ and the dot occupation $n$ are given by the following equations \cite{kaka+08}
\begin{subequations}\label{eq:rateeq}
\begin{align}
	J_{\rm L,R}(t) &= \Gamma_{\rm L,R}(t) \left[ f(\varepsilon(t)) - n(t) \right]\;,\\
		\partial_{t}n(t) &= J_{\rm L}(t) + J_{\rm R}(t)\;, \label{eq:occeom}
\end{align}
\end{subequations}
with $f(\eps)$ the Fermi distribution function describing the occupation in the reservoirs.
The pumped charge obtained in this model agrees very well with the NEGF result as can be seen in Fig.\,\ref{fig:qdpc}, which shows the current reversal in panel a ($Q$ vs frequency for $\del=-3\pi/4$) and the rectification in panel b ($Q$ vs phase shift for $\Omega=\Gamma/10$). 

\begin{figure*}[t]
	\includegraphics[width=1.6\columnwidth]{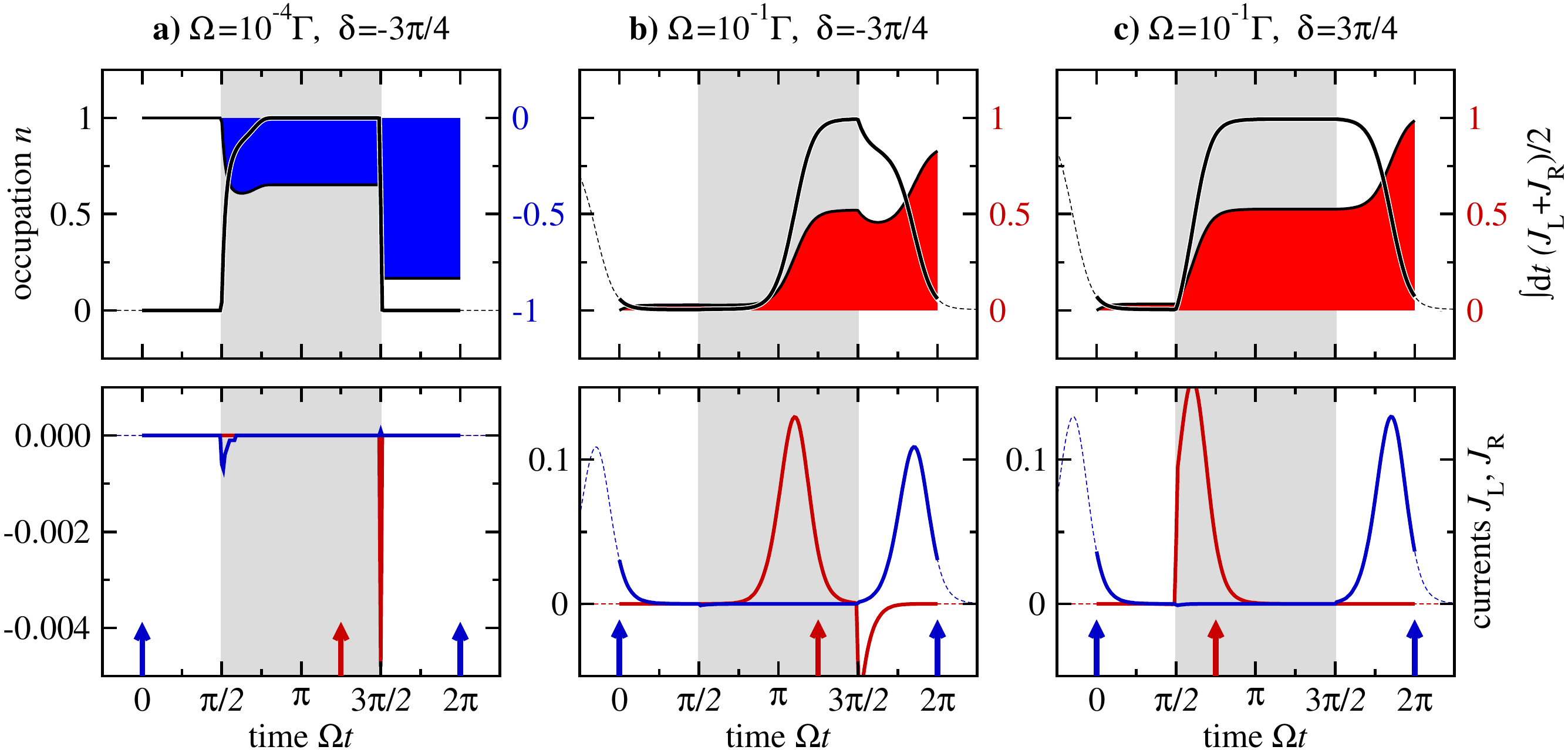}
	\caption{Time evolution within one period for three different parameters sets $(\Omega,\delta)$.
	Lower row: Currents through tunnel barriers	$J_{\rm L}(t)$ and $J_{\rm R}(t)$
	(red and blue lines).
	Upper row: Dot occupation $n(t)$ and  transferred charge $\int^{t} dt' (J_{\rm L}(t')-J_{\rm R}(t'))/2$.
	The gray area indicates the time span when the dot is charged, i.e., $\eps(t)<0$.
	Red/blue arrows mark the times when left/right couplings are maximal.
	}
	\label{fig:timeev}
\end{figure*}%
In order to get a better understanding of this surprising behavior, it is instructive
to examine the temporal evolution of the currents for slow and fast drivings, respectively. 
This is most conveniently done using a Fourier analysis.
By means of the definition \eqref{eq:Qdef} and the equation of motion \eqref{eq:rateeq} the pumped charge per period reads
\begin{align}
	Q	&= \frac{1}{2}\int^{\tau}_0 dt'\;  
			\Gamma^{(-)}(t')
			\left[ f(\varepsilon(t')) - n(t') \right]\;. \label{eq:DefQtransfer}
\end{align}
It is given in terms of the difference of the tunnel couplings $\Gamma^{(-)}$, which is
defined (along with the corresponding sum $\Gamma^{(+)}(t)$ which will be used later) as follows
\begin{align}
	\Gamma^{(\pm)}(t) \equiv{}&  \Gamma_{\rm L}(t) \pm \Gamma_{\rm R}(t) 
		={} \sum_m \Gamma^{(\pm)}_m e^{\imath m \Omega t}\;.
\end{align}
Here, $\Gamma^{(\pm)}_m = \Gamma^0_{\rm L} I_m(\eta_{\rm L}) e^{\imath m \del \pi} 
\pm \Gamma^0_{\rm R} I_m(\eta_{\rm R})$ with
$I_m$ is the modified Bessel function of the first kind of order $m$.
The last expression is the Fourier series of the tunneling-rate sum/difference. Analogously,
the occupation $n(t) = \sum_m n_m e^{\imath m \Omega t}$ and $f(\varepsilon(t))= \sum_m f_m e^{\imath m \Omega t}$ can be expanded in a Fourier series. 
For sufficiently low temperatures, $k_{\rm B} T\ll\eps_{1}$, we can replace the Fermi function by the step function $f(\eps(t))=\Theta(-\eps(t))$ and get
$f_{m}\equiv\Theta_m$ with $\Theta_{0}{=}1/2$, $\Theta_{m}{=}(-1)^{(m+1)/2}/m\pi$ for odd $m$, and $\Theta_{m}{=}0$ for even $m{\ne}0$.

Plugging these series into Eqs.\ \eqref{eq:rateeq} yields an algebraic equation for the
Fourier coefficients of the occupation, which reads in matrix-vector notation
\begin{equation}\label{eq:occsystem}
	\imath \Omega\, \mathbb{D} \cdot \bg{n} = \mathbb{G}\cdot\left[\bg{\Theta}-\bg{n}\right],
\end{equation}
with the time-derivative operator $D_{km} \equiv k\,\delta_{mk}$ and the coupling matrix $G_{km} \equiv \Gamma^{(+)}_{k-m}$. 
The components of the vectors $\bg{\Theta}$, $\bg{n}$ and $\bg{\bar{\Gamma}}^{(-)}$,
are given by the Fourier coefficients introduced above
\footnote{Notice, for $\bg{\bar{\Gamma}}$ the components are reversed,
$\bar{\Gamma}^{(\pm)}_m=\Gamma^{(\pm)}_{-m}$.}. 
It is important to notice, that neither $\mathbb{D}$ nor $\mathbb{G}$ contain the frequency $\Omega$.
This allows for a straightforward expansion in powers of $\Omega$, as we will show below. 
For the charge $Q$ in Eq.\,\eqref{eq:DefQtransfer} one needs $\bg{\Theta}-\bg{n}$, which can be easily obtained from Eq.\,\eqref{eq:occsystem} yielding
\begin{subequations}\label{eq:QpCresult}\begin{align}
Q &= \frac{\pi}{\Omega} \bg{\bar{\Gamma}}^{(-)} {\cdot}\, \left[ \bg{\Theta} - \bg{n} \right] \;,\\
\bg{\Theta} - \bg{n} &=\left[\mathbb{D} -\frac{\imath}{\Omega} \mathbb{G} \right]^{-1} \mathbb{D}\,\bg{\Theta}.
 \end{align}\end{subequations}
Note that this expression only depends on given quantities, which are either external parameters (like $\bg{\bar{\Gamma}}^{(-)}$ or $\Omega$) or trivial matrices (like $\mathbb{D}$).
With this formulation one can derive intuitive expressions for low- and high-frequency pumping. 
In order to invert the matrix in Eq.\,\eqref{eq:QpCresult} we split the matrix $\mathbb{G}/\Omega=\mathbb{G}_{0}+\mathbb{G}_{1}$ into a diagonal and an off-diagonal component
\footnote{Note that all $G_{kk}=I_{0}(\eta_{l})+I_{0}(\eta_{r})\equiv\Gamma_{0}$.}
\begin{subequations}
\begin{align}
(G_{0})_{km} &\equiv \delta_{km}G_{km}/\Omega= \delta_{km}\Gamma_{0}/\Omega \;,\\
(G_{1})_{km} &\equiv (1-\delta_{km})G_{km}/\Omega \;,
\end{align}
\end{subequations}
and use for Eq.\,(\ref{eq:QpCresult}b) the expansion
\begin{subequations}\label{eq:diffexpansion}
\begin{align}
	\bg{\Theta} - \bg{n} &= \sum_{k=0}^{\infty}\mathbb{K}_{k}\:\mathbb{D}\;\bg{\Theta}\;, \\
	\mathbb{K}_{k} &\equiv \left(\mathbb{D} {-}\imath \mathbb{G}_{0} \right)^{-1}
	\left[\mathbb{G}_{1}\left(\mathbb{D} {-}\imath \mathbb{G}_{0} \right)^{-1}\right]^{k}\;,
\end{align}
\end{subequations}
where $\mathbb{D} {-}\imath \mathbb{G}_{0}$ can be easily inverted since it is diagonal.
Equation \eqref{eq:diffexpansion} has a very intuitive interpretation.
The Fourier vector $\mathbb{D} \bg{\Theta}$ describes alternating ``$\delta$-kicks''
at times $\Omega t_{j}=(j{+}1/2)\pi$.
The expansion in terms of $\mathbb{K}_{k}$ accounts for the response of the system 
to these kicks, which is mainly characterized by the ratio $\Gamma_{0}/\Omega$.
The first term $\mathbb{K}_{0}$ in the sum is diagonal and given as
\begin{equation}\label{eq:K0}
(K_{0})_{mm} 
=\frac{m+\imath\Gamma_{0}/\Omega}{m^{2}{+}(\Gamma_{0}/\Omega)^{2}}.
\end{equation}
The Lorentzian decay in the index $m$ accounts for the exponential charging or discharging of the quantum dot. 
It is interesting to consider the following limits  
\begin{subequations}\label{eq:K0limits}
\begin{align}
\Gamma_{0}/\Omega\gg1:\quad& \mathbb{K}_{0}= \imath\frac{\Omega}{\Gamma_{0}}\mathbb{I}+\left(\frac{\Omega}{\Gamma_{0}}\right)^{2}\mathbb{D} +\ldots\\
\Gamma_{0}/\Omega\ll1:\quad& \mathbb{K}_{0}= \tilde{\mathbb{D}}^{-1}+\imath\frac{\Gamma_{0}}{\Omega} \tilde{\mathbb{D}}^{-2} +\ldots
\end{align}
\end{subequations}
with $\tilde{D}_{kk}\equiv D_{kk}$ for $k\ne0$ and $\tilde{D}_{00}\equiv-\imath\Gamma_{0}/\Omega$, which replaces $\mathbb{D}$ in order to enable the matrix inversion.
 
In the adiabatic limit ($\Omega/\Gamma_{0}\to0$) given by\ Eq.\,(\ref{eq:K0limits}a), all matrix elements of $\mathbb{K}_{0}$ are identical. 
This implies that $\bg{\Theta} - \bg{n}\propto\mathbb{D}\bg{\Theta}$, i.e., the ``$\delta$-kicks'' of the driving mentioned before also occur in the response of the system. In other words, the electron in- our outflow is much faster than the external period. Indeed this can be seen in the lower panel of Fig.\,\ref{fig:timeev}a.
Therefore, $Q$ in Eq.\,\eqref{eq:DefQtransfer} is determined by the couplings at specific times $\Gamma^{(-)}(t_{j})$. 
For negative phase shifts $-\pi{<}\delta{<}0$, as shown in Fig.\,\ref{fig:timeev}a, it is $\Gamma^{(-)}{<}0$ for the charging at $\Omega t_{0}{=}\pi/2$ and $\Gamma^{(-)}{>}0$ for the discharging at $\Omega t_{1}{=}3\pi/2$.
The opposite applies for positive shifts $0{<}\delta{<}{+}\pi$. As mentioned before, the pumping is ``peristaltic'' \cite{br98}.
The simple relation of $Q$ and the coupling differences $\Gamma^{(-)}(t_{0})$ and $\Gamma^{(-)}(t_{1})$ explains that the maximal charge $Q$ is obtained for $\delta=\pm\pi/2$ and that it vanishes for $\delta=0,\pm\pi$.  
Because of the pre-factor in Eq.\,\eqref{eq:DefQtransfer}, $Q$ becomes independent of $\Omega$ in the adiabatic case and the first non-adiabatic correction is proportional to $\Omega$.

On the other hand, in the fast driving limit ($\Omega/\Gamma_{0}\to\infty$) given by Eq.\,(\ref{eq:K0limits}b), we get $\bg{\Theta} - \bg{n}\propto\bg{\Theta}+\mbox{const}$ and the occupation $n(t)$ is constant.
Because of Eq.\,\eqref{eq:DefQtransfer} there is no transfer in this limit and $Q\propto\Omega^{-1}$. 
In between these two extrema the time scale of the exponential decay is comparable to the period of the external driving.
Thus the integral \eqref{eq:DefQtransfer} ``considers'' $\Gamma^{(-)}(t)$ over the whole period, not just at particular instants of time as in the adiabatic case.
This explains why the net current flows in the same direction as long as the peak of the left coupling occurs during charging of the dot, which is fulfilled for $|\delta|{>}\pi/2$, cf.\ gray area in Fig.\,\ref{fig:timeev}.
For $\delta{=}{-}3\pi/4$ (Fig.\,\ref{fig:timeev}b) the charging occurs in the 2nd half of the charging period,
for $\delta{=}{+}3\pi/4$ (Fig.\,\ref{fig:timeev}c) in the 1st half.
Thus, in both cases the dot is charged from the left and discharged to right since the right coupling $\Gamma_{\rm R}$ is locked to the oscillating level $\eps$. This explains the observed rectification effect.
By means of this interpretation one would expect a current reversal for phase delays $-\pi{<}\delta{<}{-}\pi/2$.
For these delays it is $\Gamma^{(-)}(t_{0}){<}0$, relevant for small $\Omega$, but $\int_{t_{0}}^{t_{1}}{\rm d}t'\,\Gamma^{(-)}(t'){>}0$, relevant for large $\Omega$, and the charging occurs either form the right or the left.
Correspondingly, it is $\Gamma^{(-)}(t_{1}){>}0$ and $\int_{t_{1}}^{t_{2}}{\rm d}t'\,\Gamma^{(-)}(t'){<}0$ and the dot is discharged to the opposite direction.
Figure \ref{fig:qdpc3D} indeed shows this behavior in the predicted range of phase delays $\delta$.

Finally, to show that the current reversal is not specific to our driving scheme, we turn
to the case of purely harmonic driving, i.e., $\Gamma(t)=\Gamma_{0}+\Gamma_{1}\cos(\Omega t{-}\delta)$. More importantly, the basic mechanism of the current reversal can be understood analytically in this case.
Harmonic driving at a frequency $\Omega$ is characterized by having three Fourier components
$\mathbf{\Gamma}_{\rm L}=\{\Gamma_{1}\e{+\imath\delta},\Gamma_{0},\Gamma_{1}\e{-\imath\delta}\}$
and $\mathbf{\Gamma}_{\rm R}=\{\Gamma_{1},\Gamma_{0},\Gamma_{1}\}$, where
$\Gamma_{0}>\Gamma_{1}$ guarantees positive couplings 
and $\delta$ is the time shift of the left and right coupling.
If the calculation in Eq.\,\eqref{eq:diffexpansion} is restricted to $\mathbb{K}_{0}$ 
only three  
Fourier components of the step function are needed, which are $\mathbf{\Theta}=\{-1/\pi,1/2,-1/\pi\}$.
Using these expressions in Eqs.\,\eqref{eq:QpCresult}--\eqref{eq:K0} one gets
\begin{equation}\label{eq:Qharm}
	Q=\frac{\Gamma_{1}}{\Gamma_{0}{}^{2}{+}\Omega^{2}}\left[\Omega\:(1{-}\cos\delta)+\Gamma_{0}\:\sin\delta\right].
\end{equation}
This simple expression contains all the basic features for periodic pumping including the current reversal. Moreover, it allows us to analyze the respective regimes in
detail. For $\Omega\ll\Gamma_{0}$ (adiabatic limit) one obtains from Eq.\ \eqref{eq:Qharm}
$Q=(\Gamma_{1}/\Gamma_{0}) \sin\delta$. As discussed earlier, the sign of $Q$ depends
on the order of the ``door openings''. Optimal transfer is attained for 
$\Gamma_{1}=\Gamma_{0}$. In the opposite limit, $\Omega\gg\Gamma_{0}$, one finds
$Q=(\Gamma_{1}/\Omega)(1{-}\cos\delta)$. Most interestingly and in contrast to the
adiabatic case, in this limit the sign of $Q$ is independent of the phase shift $\delta$, 
which is the non-adiabatic rectification effect. Consequently, for $\delta<0$ one gets
negative Q in the adiabatic and positive Q in the non-adiabatic limit: the 
average current can be reversed by tuning the driving frequency. 
These findings are confirmed by Fig.\ \ref{fig:qdpc}, where we compare the harmonic driving to 
the scenario considered initially [Eqs.\ \eqref{eq:driving}]. Qualitatively, the behavior
of $Q$ is quite similar in both cases, which underlines the robustness of the 
discussed effects.

In summary, we have studied the influence of non-adiabatic driving on the charge
pumping through a quantum dot in the Coulomb blockade regime. Our numerical 
calculations, based on a NEGF method, showed that the average pumped current 
can be reversed by sweeping the
driving frequency. The origin of this effect was found to be the qualitatively
different response to slow and fast driving, rendering the difference of the
left and right tunneling rates matter only at specific instants of time 
(adiabatic case) or during a time-interval (non-adiabatic case).
By means of a description with rate equations, we derived for the case of harmonic driving
an analytical expression for the transferred charge per cycle, which confirms our analysis. 
Furthermore this shows that the observed effects are generic and quite robust with respect to the specific form of the external driving.
Therefore they could be useful for realizing frequency filters or frequency-selected switches.

\emph{Acknowledgments.} We thank Zach Walters for proofreading the manuscript.


\end{document}